\documentclass[aps,pre,twocolumn,floatfix,superscriptaddress,showpacs]{revtex4-1}
\usepackage{graphicx}
\begin{document}
\setcounter{page}{1}
\title[]{Spectral dimensions of hierarchical scale-free networks with shortcuts}
\author{S.~\surname{Hwang}}
\author{C.-K~\surname{Yun}}
\affiliation{Department of Physics and Astronomy, Seoul National University, Seoul 151-747, Korea}
\author{D.-S.~\surname{Lee}}
\email{deoksun.lee@inha.ac.kr}
\affiliation{Department of Natural Medical Sciences and Department of Physics, Inha University,
Incheon 402-751, Korea}
\author{B.~\surname{Kahng}}
\email{bkahng@snu.ac.kr}
\affiliation{Department of Physics and Astronomy, Seoul National University,
Seoul 151-747, Korea}
\affiliation{School of Physics, Korea Institute for Advanced Study, Seoul
130-722, Korea}
\author{D.~\surname{Kim}}
\affiliation{Department of Physics and Astronomy, Seoul National University, Seoul 151-747, Korea}

\begin{abstract}
The spectral dimension has been widely used to understand transport properties on
regular and fractal lattices. Nevertheless, it has been little studied for complex
networks such as scale-free and small world networks. Here we study the spectral
dimension and the return-to-origin probability of random walks on hierarchical
scale-free networks, which can be either fractals or non-fractals depending
on the weight of shortcuts. Applying the renormalization group (RG) approach to
the Gaussian model, we obtain the spectral dimension exactly.
While the spectral dimension varies between $1$ and $2$ for the fractal case,
it remains at $2$, independent of the variation of network structure
for the non-fractal case. The crossover behavior between the two cases
is studied through the RG flow analysis. The analytic results are confirmed
by simulation results and their implications for
the architecture of complex systems are discussed.
\end{abstract}


\maketitle

\section{INTRODUCTION}

The problem of random walks (RWs) on complex networks have attracted much attention
as a model to study diffusion processes on complex systems such as fad or disease
spreading over social networks, data packets transport in the Internet, data mining
on the web, and so on \cite{hughesbook,havlinbook,rednerbook,albert02,dorogovtsev03a,newman03}.
Such importance in theoretical and application aspects has led to the extensive studies on the
problem \cite{sven01,noh04,masuda04,sood05,bollt05,condamin07,baronchelli08,zhang09,tejedor09},
but it still remains unclear how RW properties depend on network structure.
In this paper, we investigate the effect of shortcut links on RW motion.
Shortcuts may be classified into two types: Long-range shortcuts, utilized
in a skeleton or superhighways, and local shortcuts, complementary to the skeleton or
local roads \cite{dhkim04,goh06,wu06}. It was recently shown systematically \cite{rozenfeld10}
that long-range shortcuts can change the type of networks from fractal to
non-fractal. The fractal (non-fractal) network is the one in which the mean
separation between two nodes scales with system size in a power-law
(logarithmic) manner. We are particularly interested in how
RW properties are changed as long-range shortcuts are added and thereby
the network changes from a fractal to a non-fractal network.

The Gaussian model is useful to study RW problems analytically~\cite{gaussian}, which
we use to understand the RW motion on hierarchical scale-free
networks \cite{berker79,berker06,rozenfeld07a,rozenfeld07b}.
In this artificial network, we can control the degree distribution and
the link weight of shortcuts. We consider RWs on this hierarchical network, and
obtain the exact solution of the return-to-origin (RTO) probability
by applying the renormalization group (RG) approach to the Gaussian model.
The RTO probability is related to the free energy of the Gaussian model and
the spectral density function of the Laplacian matrix, all of which are
characterized by the spectral dimension $d_s$ \cite{hughesbook,gaussian}.
The hierarchical network is recursively organized, so that one can obtain
analytically the spectral density function by using the decimation method
of the RG transformation. We find that when the network structure is
a fractal, the spectral dimension $d_s$ varies in the range
$1 < d_s \leq 2$; however, it is fixed at $d_s=2$ for the non-fractal networks.
These analytic results are confirmed by numerical simulations.
Moreover, we perform the numerical simulations of RWs on
various real-world networks and check the relationship between
the network fractality and the spectral dimension.

The paper is organized as follows. In Section~\ref{sec:modelformalism},
the hierarchical network model and the formalism to derive
the spectral density function of the Laplacian matrix from the
Gaussian model is briefly introduced along with their
relation to the RTO probability of RWs.
In Section~\ref{sec:rg}, we apply the RG approach to the Gaussian model
and obtain the spectral density function explicitly.
In Section~\ref{sec:simulation}, numerical results
of the RTO probability are presented and compared
with the analytic solutions. We summarize our findings and
discuss their implications in Section~\ref{sec:summary}.

\section{Model and formalism}
\label{sec:modelformalism}

\begin{figure}
\includegraphics[width=8.5cm]{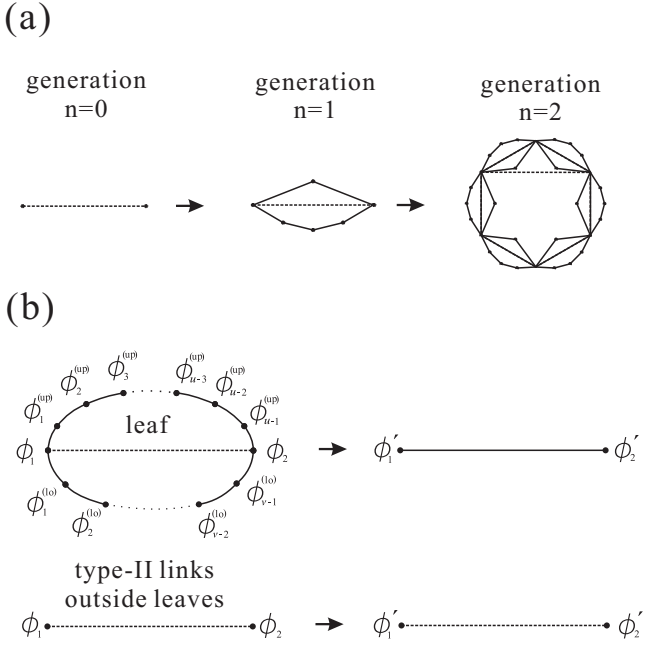}
\caption{Weighted Flower (WF) network and the renormalization of
  the Gaussian model.
  (a)  Construction of the WF network with $u=2$ and $v=4$
    from generation $n=0$ to $n=2$. Type-I links (solid line)
   and type-II links (dotted line) have link weight $1$ and $p$, respectively.
  (b) Renormalization of a leaf and a type-II link. A leaf has two chains
  and one type-II link. The upper chain has
  $u-1$ $\phi$ variables and the lower chain has $v-1$ $\phi$ variables.
  After a RG transformation, each leaf is
  replaced by a type-I link. A type-II link outside the leaves remains
  after a RG transformation.}
\label{fig:model}
\end{figure}

The network we use in the paper is a modified version of an existing
hierarchical network model \cite{berker79,berker06,rozenfeld07a,rozenfeld07b},
in which shortcuts are assigned weights.  We call this modified model
the {\it weighted flower} (WF) network.
The WF network contains two types of links, type I and II, which evolve differently.
The WF network is constructed iteratively as follows.
We start from a single type-I link between two nodes, which is the 0-th generation.
At the next generation, two chains are added between the two nodes at both ends of
the type-I link. On one chain, $u-1$ nodes are located and thereby $u$ links are placed.
On the other chain, $v-1$ nodes are located and thereby $v$ links are placed.
Those $u+v$ links are assigned as type-I links, and the type of the seed link
is changed to type II. These steps - adding nodes and links and assigning link
types - are performed at every generation. Every link is assigned weight: $1$
for each type-I link and $p$ for each type-II link. The link weights will be
used for the transition rate of RWers. The growth of the WF network with
$u=2$ and $v=4$ is depicted in Fig.~\ref{fig:model} (a).
We consider only the case $u\geq 2$ and $v\geq 2$ throughout this paper.
The type-II links play a role as shortcuts in the system  and
the parameter $p$ represents the transition rate across a type-II link.
We point out that in the previous model of the hierarchical network
\cite{berker79,berker06,rozenfeld07a,rozenfeld07b},
the parameter $p$ was used as the occupation probability of
the type-II links, but here, it is used as the weight of the type-II links.
Thus, the two network models are equivalent if $p=0$ or $p=1$.
Otherwise, they are different.

\begin{table}
\begin{tabular}{lc}
\hline
number of type-I links& $L_n^{\rm (I)}= (u+v)^n$ \\
number of type-II links &$L_n^{\rm (II)}= {(u+v)^n-1\over u+v-1}$\\
number of nodes & $N_n={u+v-2\over u+v-1} (u+v)^n + {u+v\over u+v-1}$\\
type-I degree & $k_{\ell}^{\rm (I)}=2^{n-n_\ell+1}$\\
type-II degree & $k_{\ell}^{\rm (II)}=p 2^{n-n_\ell+1}-2 p$\\
node degree & $k_{\ell}= (1+p) 2^{n-n_\ell+2} - 2p$\\
degree distribution & $p(k)\sim k^{-\gamma}$ with $\gamma = 1+{\log (u+v)\over \log 2}$\\
\hline
\end{tabular}
\caption{Basic properties of the WF network at generation $n$. Here $\ell$ is
the node index and $n_\ell$ denotes the birth generation of node $\ell$. The degree of
a node is defined here as the sum of the weights of connected links.}
\label{table:model}
\end{table}

The link weights of the WF network at the $n$-th generation are
represented in a symmetric matrix form
\begin{equation}
W_{\ell q} = \left\{
\begin{array}{ll}
1~~~ &~~~ (\ell q)\in E_n^{\rm (I)},\\
p~~~ &~~~ (\ell q)\in E_n^{\rm (II)},\\
0~~~ &~~~ {\rm otherwise},
\end{array}
\right.
\label{eq:wlq}
\end{equation}
where $E_{n}^{\rm (I)}$ and $E_n^{\rm (II)}$ denote
the set of type-I and type-II links at the $n$-th generation, respectively,
and $(\ell q)$ represents the link connecting node $\ell$ and $q$.
Basic structural properties of the WF networks can be
obtained analytically~\cite{berker06,rozenfeld07a,rozenfeld07b} and
are summarized in Table~\ref{table:model}. Here we defined the
degree $k_{\ell}$ of node $\ell$ as $k_{\ell}=\sum_q W_{q \ell}$.
The distance between two nodes was defined as the
length of the minimum-cost path with the link cost $c_{\ell q}$ given by
a decreasing function of $W_{\ell q}$ satisfying $c_{\ell q}\to \infty$ as
$W_{\ell q}\to 0$~\cite{noh02}.
While the degree exponent depends on $u$ and $v$
but remains the same under the variation of the weight $p$,
the mean distance between nodes depends on $p$.
When $p=0$, the type-II links are not considered in
determining the minimum-cost paths. Then
the mean distance $D$ between nodes in the WF networks
is related to the total number of nodes as
~\cite{berker06,rozenfeld07a,rozenfeld07b}
\begin{equation}
D\sim N^{1/d_f} \ {\rm with} \  d_f = {\log(u+v)\over \log(\min\{u,v\})}.
\label{eq:Nlp0}
\end{equation}
Here $d_f$ is the fractal dimension~\cite{song06} of the WF networks with $p=0$.
When $0<p\leq 1$, shortcuts may participate in the minimum-cost paths and
the mean distance is proportional to the
number of nodes logarithmically~\cite{berker06,shortestpath}
\begin{equation}
D\sim \log N.
\label{eq:Nlp1}
\end{equation}

Let us consider a RWer located initially at node
$\ell_0$ and going around in the network with the transition probability
$W_{\ell q}/\sum_{\ell^{\prime}} W_{\ell^{\prime} q}$ from node $q$ to $\ell$.
Then the probability $P_{\ell \ell_0}(t)$ to find the RWer at node $\ell$
after time $t$ is given by $(I-L)^t_{\ell \ell_0}$,
where $I$ is the identity matrix and $L$ is the Laplacian matrix defined as
$L_{\ell q}=\delta_{\ell q} - W_{\ell q}/k_q$ with
$k_q=\sum_{\ell} W_{\ell q}$.

The RTO probability $P_o(t)=N^{-1}\sum_{\ell} P_{\ell \ell}(t)$ is determined by the
eigenvalues $\mu_i$ ($i=0,1,2,\ldots, N-1)$ of the Laplacian matrix $L$ as
\begin{eqnarray}
P_o(t) &=& {1\over N}{\rm Tr} \, (I-L)^t = {1\over N}
\sum_{i=0}^{N-1} (1-\mu_i)^t\nonumber\\
&\sim& \int_0^\infty d\mu\, \rho(\mu)\, e^{-\mu t}+
(-1)^t\int_0^\infty d\tilde{\mu} \, \rho(2-\tilde{\mu}) \,
e^{-\tilde{\mu}t}, \nonumber \\
\label{eq:PRTO}
\end{eqnarray}
where we introduced the spectral density function
$\rho(\mu)\equiv N^{-1}\sum_{i=0}^{N-1} \delta(\mu-\mu_i)$
in the thermodynamic limit $N\to\infty$~\cite{perron}.
For large $t$, the term $(1-\mu_i)^t$ in Eq.~(\ref{eq:PRTO})
is dominant as $\mu_i\to 0$ and $\mu_i\to 2$.
The contribution from $\mu_i\simeq 2$, however, cannot be larger
than that from $\mu_i\simeq 0$ since $P_o(t)\geq 0$, but
can cause an oscillatory behavior in $P_o(t)$ depending on whether $t$ is even or odd.
For example, in $d$-dimensional regular lattices, the eigenvalues are
symmetrically distributed about $\mu=1$, and $\rho(\mu)=\rho(2-\mu)\sim \mu^{d/2-1}$
for small $\mu$. Then, it follows from Eq.~(\ref{eq:PRTO}) that
$P_o(t)\sim t^{-d/2}$ when $t$ is even and $P_o(t)=0$ when $t$ is odd.

For fractal lattices, the spectral density function and the RTO probability are
characterized by the spectral dimension $d_s$ as~\cite{havlinbook}:
\begin{eqnarray}
\rho(\mu)&\sim& \mu^{d_s/2-1} \ {\rm for} \ \mu \to 0, \ {\rm and} \nonumber\\
P_o(t)&\sim& t^{-d_s/2} \ {\rm for} \ t \to \infty.
\label{eq:dsdef}
\end{eqnarray}
We use these formulae in further discussion later.

The mean first passage time $T$ between two nodes is related to the
eigenvalues of the Laplacian matrix via $T=\sum_i \mu_i^{-1}$~\cite{hughesbook},
which scales with system size $N$ as $T \sim N^{2/d_s}$ for
$d_s < 2$ and $T\sim N$ for $d_s > 2$~\cite{bollt05}.
For the fractal WF network with
$p=0$ in the $n$-th generation, it was shown that
$T \sim (uv)^n \sim N^{\log(uv)/\log(u+v)}$ based on the scaling behavior
of the mean first passage time between two hub nodes \cite{rozenfeld07a}.
Thus, it follows that $d_s=2\log(u+v)/\log(uv)$ for the case $p=0$.
On the other hand, the mean first passage time for a non-fractal WF network
with $u=1$, $v=2$ and $p=0$ was calculated, which scales as
$T\sim 3^n\sim N$ \cite{bollt05,rozenfeld07a}.
Thus, the spectral dimension is $d_s=2$ in this case.
These results suggest that the spectral dimension can be different depending
on whether a network is fractal or non-fractal. Motivated by these previous
works, we study the spectral dimension of the hierarchical network as it
crosses over from a fractal to a non-fractal by varying
the parameter $p$ in the range $0\leq p\leq 1$.

To derive the spectral density function of the matrix $L$,
we use the Gaussian model of which the partition function is given as
\begin{equation}
Z(\mu)=  \int \mathcal{D} \phi \, \exp\left[
i \sum_{\ell,q} \phi_\ell H_{\ell q}(\mu) \phi_q\right] =  \sqrt{\frac{(i \pi)^{N}}{\det H}},
\label{eq:PF}
\end{equation}
where $N$ is the number of nodes, $\mathcal{D}\phi=\prod_{\ell=1}^N d\phi_\ell$,
and $H=\mu I-\tilde{L}$.
$\tilde{L}$ is a similarity-transformed Laplacian matrix given in
Appendix~\ref{seca:laplacian}.
$\mu$ is a parameter introduced in the Gaussian model.
The spectral density function $\rho(\mu)$ is derived from
the free energy $f(\mu)\equiv -\lim_{N\to\infty}N^{-1}\log Z(\mu)$ as
\begin{equation}
\rho(\mu)=-{2\over \pi}{\rm Im} {\partial f\over \partial \mu},
\label{eq:rhof}
\end{equation}
where $\mu$ is assumed to possess a positive infinitesimal imaginary part.

\section{Renormalization group approach to Gaussian model}
\label{sec:rg}

\subsection{Fixed points and RG flow}

The parameter $\mu$ in Eq.~(\ref{eq:PF}) is renormalized differently
for the two types of links in the WF networks. Thus,
we first rescale $\phi_\ell$ as $\phi_\ell/\sqrt{k_{\ell}}$ in Eq.~(\ref{eq:PF})
and introduce distinct parameters $\mu_1$ and $\mu_2$ for the respective link types.
The Hamiltonian then takes the form
$H_{\ell q}=[(\mu_1-1)k_{\ell}^{\rm (I)}+(\mu_2-1) k_{\ell}^{\rm (II)}]\delta_{\ell q}+W_{\ell q}$,
where $W_{\ell q}$ is given in Eq.~(\ref{eq:wlq}). The partition function
in the n-th generation is written as
\begin{eqnarray}
&&Z_n (\mu_1,\mu_2,p) \nonumber\\
&&=\int \mathcal{D}\phi \prod_{(\ell q)\in E_n^{\rm (I)}} z_n^{\rm (I)}(\phi_\ell,\phi_q)
\prod_{(\ell q)\in E_n^{\rm (II)}} z_n^{\rm (II)}(\phi_\ell,\phi_q), \nonumber\\
&&{\rm where}\nonumber \\
&&z_n^{\rm (I)}(\phi_\ell,\phi_q)= \exp\left(-i(1-\mu_1)(\phi_\ell^2 +
      \phi_q^2) +2i\phi_\ell\phi_q\right),\nonumber\\
&&{\rm and}\nonumber \\
&&z_n^{\rm (II)}(\phi_\ell,\phi_q)=\exp\left(-i(1-\mu_2)p(\phi_\ell^2 +\phi_q^2)+2i p \phi_\ell \phi_q \right).\nonumber \\
\label{eq:Lp1}
\end{eqnarray}
Note that $\mu_1=\mu_2=\mu$ initially.
The RG transformations for the shortcut weight $p$, and the parameters $\mu_1$, and $\mu_2$
are carried out to obtain the spectral dimension $d_s$.
As we show later, $\mu_2$ is invariant under the RG transformation and we
investigate the fixed points and the RG flow in the $(\mu_1,p)$ plane.
If there is only one fixed point at $p=0$, then the spectral dimension $d_s$
will be the same regardless of $p$. On the other hand, more than one fixed points
will indicate a crossover or a transition of $d_s$ depending on $p$.

The WF network in the $n$-th generation contains $L_{n-1}^{\rm (I)}$ leaves
and $L_{n-1}^{\rm (II)}$ links, and  the partition function may be written as
\begin{eqnarray}
Z_n = \int \mathcal{D}\phi' \prod_{(\ell'q')
  \in E_{n-1}^{\rm (I)}} Z_{\rm leaf}(\phi_\ell',\phi_q')
\prod_{(\ell'q') \in E_{n-1}^{\rm (II)}}
z_{n-1}^{\rm (II)}(\phi_\ell',\phi_q'). \nonumber
\end{eqnarray}
Here, each leaf contains $u+v$ nodes, among which $u+v-2$
are the youngest nodes, and the two remaining old nodes are
connected by a type-II link as shown in Fig.~\ref{fig:model}(b).
Integrating out the $\phi$ variables on the youngest nodes, we find
that  $Z_{\rm leaf}(\phi_1', \phi_2')$ takes a similar form to
$z_{n-1}^{\rm (I)}(\phi_1',\phi_2')$ as
\begin{eqnarray}
&& Z_{\rm leaf} (\phi_1',\phi_2')  =(\pi i/2)^{(u+v-2)/2} G(\mu_1)^{-1/2}\times \nonumber\\
&&\exp\left[-i\left\{
2(1-\mu_1) + p(1-\mu_2) -h_1(\mu_1)\right\}(\phi_1^{'2}+\phi_2^{'2})\right. \nonumber\\
&&\left.+2i \left\{p + h_2(\mu_1)\right\} \phi_1'\phi_2' \right],
\label{eq:rgleaf}
\end{eqnarray}
where
\begin{eqnarray}
c(\ell,a,b) &=& 1-a -\cos(\ell \pi/b),\nonumber\\
G(\mu_1) &=&
\prod_{\ell=1}^{u-1} c(l,\mu_1,u) \prod_{q=1}^{v-1} c(q,\mu_1,v), \nonumber\\
h_1(\mu_1) &=& {1\over u}\sum_{\ell=1}^{u-1} {\sin^2(\ell\pi/u) \over c(\ell,\mu_1,u)}
+ {1\over v}\sum_{q=1}^{v-1} {\sin^2(q\pi/v) \over c(q,\mu_1,v)},\nonumber\\
{\rm and} \nonumber \\
h_2(\mu_1) &=& {1\over u}\sum_{\ell=1}^{u-1} {(-1)^{\ell-1}\sin^2(\ell\pi/u)
  \over c(\ell,\mu_1,u)}\nonumber\\
&+& {1\over v}\sum_{q=1}^{v-1} {(-1)^{q-1}\sin^2(q\pi/v) \over c(q,\mu_1,v)}.
\label{eq:ch1h2}
\end{eqnarray}
The derivation of Eq.~(\ref{eq:rgleaf}) and
the properties of $G(\mu_1), h_1(\mu_1)$, and $h_2(\mu_1)$ for small $\mu_1$
are given in Appendices ~\ref{seca:derivation} and ~\ref{seca:expansion}.
In order for $Z_{\rm leaf}(\phi_1',\phi_2')$ in Eq.~(\ref{eq:rgleaf}) to
take the same functional form as $z_{n-1}^{\rm (I)}(\phi_1',\phi_2')$,
all the $\phi$ variables should be rescaled as
$\phi'/\sqrt{p + h_2(\mu_1)}$ so that the coefficient of $\phi_1'\phi_2'$
be unity. With the introduction of $\mu_1'$, $\mu_2'$, and $p'$ given by
\begin{eqnarray}
1-\mu_1^\prime &=& {2(1-\mu_1) + p (1-\mu_2) -h_1 (\mu_1)\over p + h_2(\mu_1)}, \nonumber\\
\mu_2^\prime &=& \mu_2,\nonumber\\
p^\prime &=& {p \over p + h_2 (\mu_1)},
\label{eq:RGpara}
\end{eqnarray}
the partition function $Z_n$ is related to $Z_{n-1}$ as
\begin{equation}
Z_n(\mu_1,\mu_2,p)= \exp[-N_n g(\mu_1,p)] Z_{n-1}[\mu_1',\mu_2{'},p'],
\label{eq:RGZ}
\end{equation}
where the function $g(\mu_1,p)$ is defined as
\begin{equation}
g(\mu_1,p)={L_{n-1}^{\rm (I)}\over 2 N_n}\log[G(\mu_1)]
+{N_{n-1}\over 2 N_n}\log\left[p + h_2(\mu_1)\right].
\label{eq:gmu}
\end{equation}
As this RG transformation is repeated, the parameters $\mu_1$ and $p$
are renormalized successively according to Eq.~(\ref{eq:RGpara}).
The parameter $\mu_2$ remains at its initial value $\mu$.
On the other hand, the
parameter $\mu_1$ is renormalized in a non-trivial way.

For small $\mu_1$, one finds two fixed points in the $(\mu_1,p)$
plane located at
\begin{equation}
(\mu_1^*, p^*)=(0,0) ~~~ {\rm and} ~~~  (0,1-u^{-1}-v^{-1}).
\label{eq:fp}
\end{equation}
These two fixed points meet when $(u,v)=(2,2)$.
Let us first consider the case $(u,v)\ne (2,2)$.
Near the fixed point $(\mu_1^*=0,p^*=0)$,
the RG equation, (\ref{eq:RGpara}), is linearized as
\begin{equation}
\mu_1^{\prime} = uv \mu_1
 ~~~{\rm and}  ~~~
  p^{\prime}={uv \over u+v}p,
\end{equation}
and therefore, after $\tau$ RG transformations,  $\mu_1$ and $p$
take the following values:
\begin{equation}
\mu_1^{(\tau)} = (uv)^\tau \mu
  ~~~ {\rm and} ~~~
p^{(\tau)}=\left({uv\over u+v}\right)^\tau p.
\label{eq:RG1}
\end{equation}
Since $uv$ and $uv/(u+v)$ are not smaller than $1$ for $u\geq 2$
and $v\geq 2$, the fixed point $(0,0)$ is unstable.
Around the other fixed point ($\mu_1^*=0,p^*=1-u^{-1}-v^{-1})$,
the linearized RG equation is written as
\begin{eqnarray}
\left(
    \begin{array}{l}
    \mu_1^\prime\\
    p^\prime - p^*
    \end{array}
\right)
&=&R
\left(
    \begin{array}{l}
    \mu_1 \\
    p - p^*
    \end{array}
\right)
+\mu
\left(
    \begin{array}{l}
    1-u^{-1}-v^{-1} \\
    0
    \end{array}
\right), \nonumber
\end{eqnarray}
where
\begin{eqnarray}
R &=&
\left(
    \begin{array}{cc}
    u+v  & 0 \\
    -{(uv-u-v)(uv-1)(u+v)\over 3(uv)^2} & {u+v \over uv}
    \end{array}
\right).
\label{eq:RGlinear}
\end{eqnarray}
The matrix $R$ has two eigenvalues $u+v$ and $(u+v)/(uv)$
and the parameters after $\tau$ RG transformations are given by
\begin{eqnarray}
\mu_1^{(\tau)}  &=& \mu\left[
{(u+v) (uv-1)\over uv (u+v-1)}\{ (u+v)^\tau -1\} +1
\right],
\nonumber\\
p^{(\tau)}-p^* &=&
 \mu {(u+v) (uv-1) (uv-u-v)\over 3(uv)^2 (u+v-1)} \{1-(u+v)^\tau\} \nonumber\\
&+&\left(p-p^*\right) \left({u+v\over uv}\right)^\tau.
\label{eq:RG2}
\end{eqnarray}

If $(u,v)=(2,2)$, there is only one fixed point at $(\mu_1^*=0,p^*=0)$ for
small $\mu_1$. The renormalized value $p^{(\tau)}$ can be calculated
from Eq.~(\ref{eq:RGpara}) as
\begin{equation}
p^{(\tau)} = {1\over \tau+p^{-1}}.
\label{eq:j2uv22}
\end{equation}
Note that $p^{(\tau)}$ is not as small as $\mu$.
Inserting Eq.(\ref{eq:j2uv22}) into Eq.(\ref{eq:RGpara}),  a
recursive relation for $\mu_1$ is obtained,
which can be solved to give
\begin{equation}
\mu_1^{(\tau)} = {(1+3p^{-1})4^{\tau}-1 \over 3(\tau+p^{-1})} \mu.
\label{eq:RG3}
\end{equation}

\begin{figure}
\includegraphics[width=6.5cm]{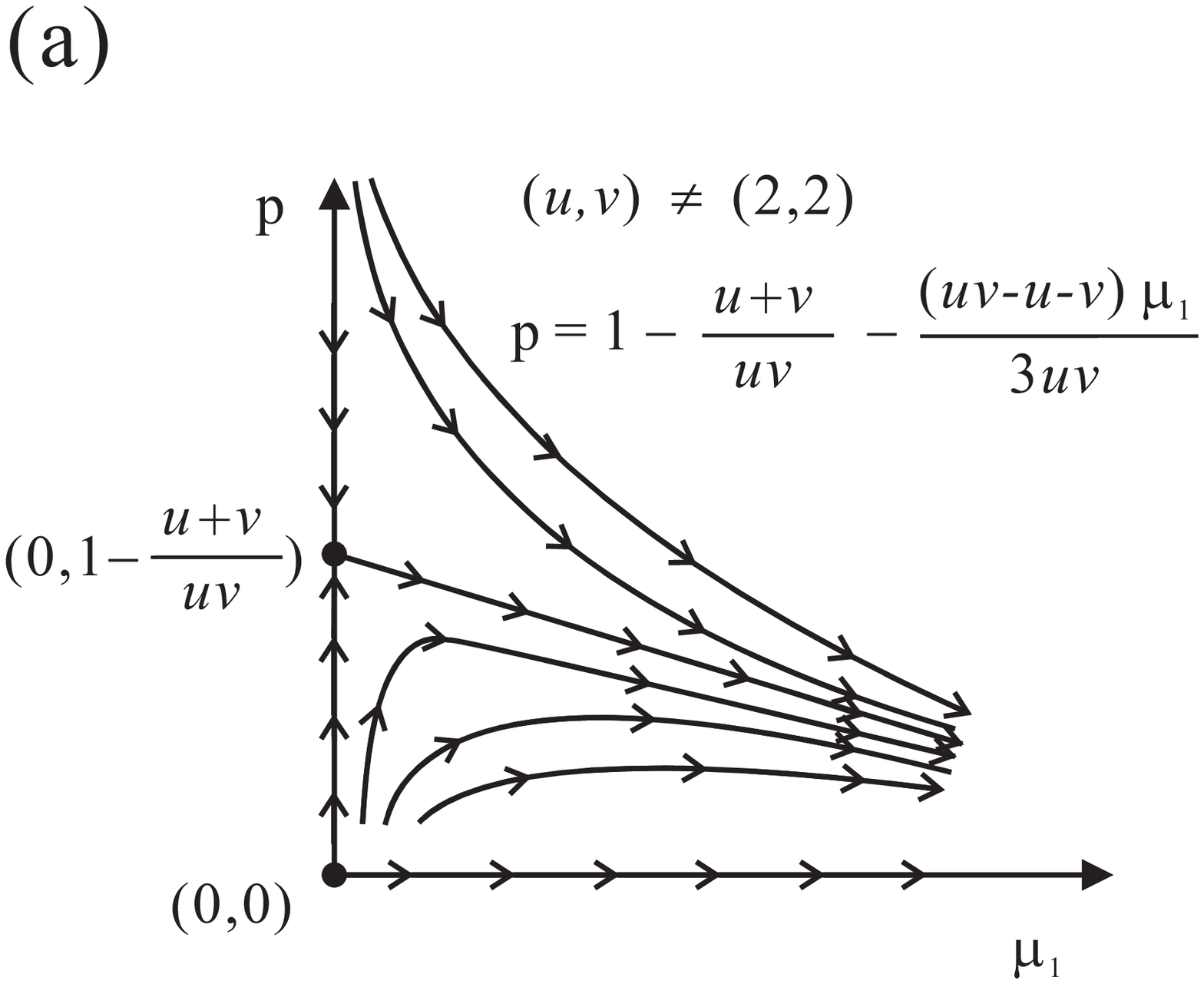}
\includegraphics[width=6.0cm]{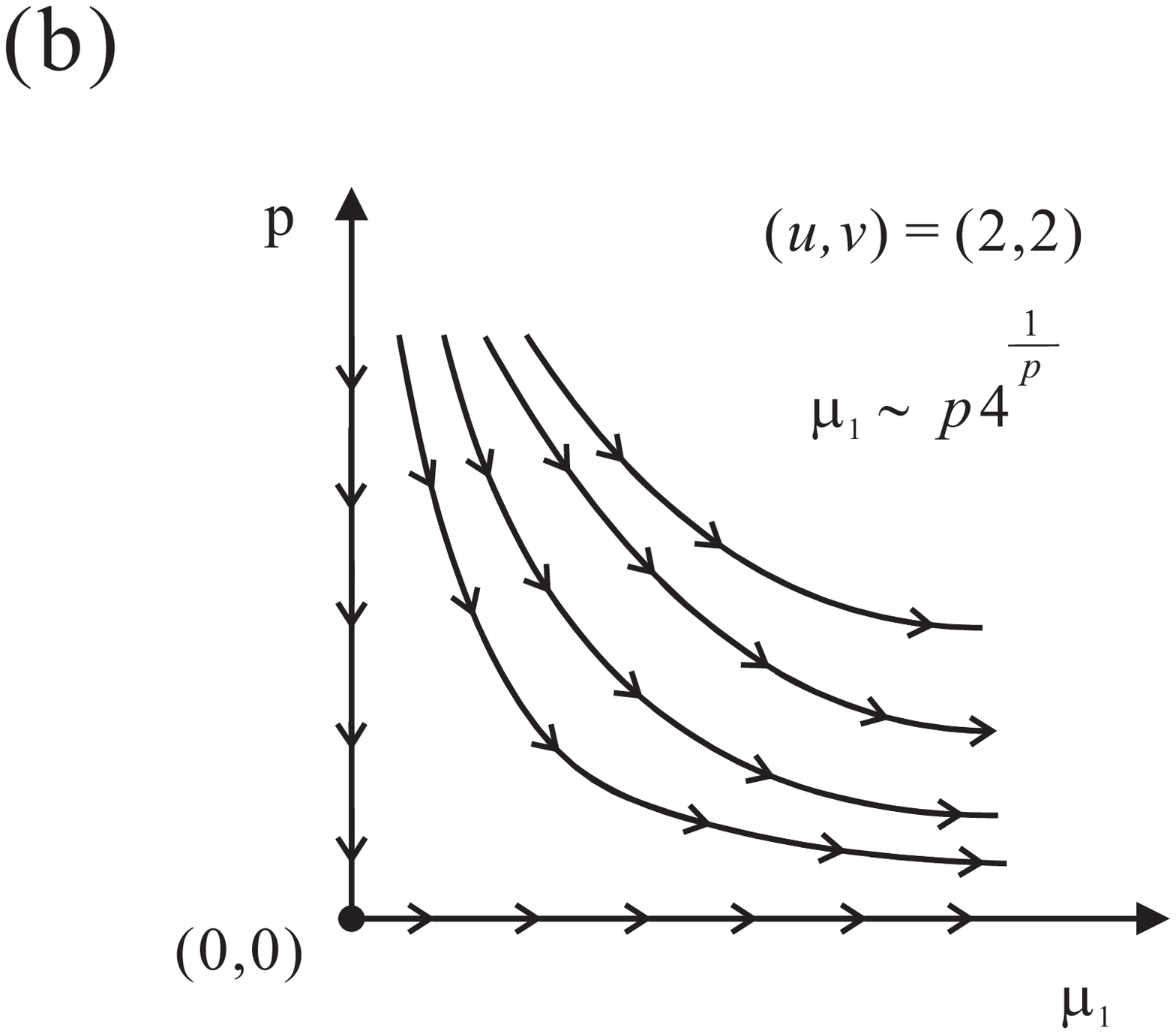}
\caption{RG flow in the $(\mu_1,p)$ parameter plane.
  (a) RG flow for $(u,v)\ne (2,2)$. There are two fixed points in the
    small-$\mu_1$ region: $(0,0)$ and $(0,1-u^{-1}-v^{-1})$. If
    $p=0$ initially, the flow is outward along the $\mu_1$ axis.
   When $p\ne 0$ initially, it is first attracted towards $(0,1-u^{-1}-v^{-1})$ and
   then repelled along the line $p=1-(u+v)/(uv)-\mu_1 (uv-u-v)/(3uv)$.
 (b) RG flow for $(u,v)=(2,2)$. There is a fixed point at the origin
 and the flow follows a curve $\mu_1 \sim p4^{1/p}$.}
\label{fig:rg}
\end{figure}

To understand the renormalization of $\mu_1$ and $p$,
which depends on their initial values as well as the
the network model parameters $u$ and $v$,
we consider the following three cases:
\begin{eqnarray}
&{\rm (I)}~~~ &  p=0, \nonumber \\
&{\rm (II)}~~~ & p\ne 0 ~~{\rm and}~~ (u,v) \ne (2,2), \nonumber \\
&{\rm (III)}~~~ &p\ne 0 ~~{\rm and}~~ (u,v)=(2,2). \nonumber
\end{eqnarray}
(I) If $p=0$ initially, $p$ remains at zero and $\mu_1$ increases
according to Eq.~(\ref{eq:RG1}).
(II) If $p > 0$ initially and $(u,v)\ne (2,2)$, both $\mu_1$ and $p$ increase
by Eq.~(\ref{eq:RG1}) and the RG flow runs outward from $(0,0)$ along the curve
$p\sim \mu_1^{1-\log(u+v)/\log(uv)}$ as obtained from Eq.~(\ref{eq:RG1}).
The RG flow then runs towards the other fixed point
$(0,1-u^{-1}-v^{-1})$, driven by one eigenvalue $(u+v)/(uv)<1$
of $R$ in Eq.~(\ref{eq:RGlinear}). Once
$p$ gets sufficiently close to $p^*$, the RG flow is bent outward
along the line $p-p^* = -\mu_1  (uv - (u+v))/(3uv)$,
parallel to the other eigenvector of $R$ (see Fig.~\ref{fig:rg} (a)).
(III) When $(u,v)=(2,2)$, the two fixed points coincide each other.
The RG flow along the $p$ axis runs quite slowly as the corresponding eigenvalue of
$R$ is $1$. If $p=0$, $\mu_1$ and $p$ are renormalized in the same way as in (I).
If $p\ne 0$, the RG flow runs along the curve $\mu_1\sim p 4^{1/p}$ as obtained from Eq.~(\ref{eq:RG3}),
which is depicted in Fig.~\ref{fig:rg}(b).

\subsection{Spectral density function and spectral dimension}

The free energy per node in the thermodynamic limit
$f(\mu) = -\lim_{n\to\infty} N_n^{-1}\log Z_n(\mu,\mu,p)$ can be
obtained by using $g(\mu_1,p)$ in Eq.(\ref{eq:gmu}) as ~\cite{cardybook}
\begin{equation}
f (\mu)= \sum_{\tau=0}^\infty {g(\mu_1^{(\tau)},p^{(\tau)})\over (u+v)^\tau},
\label{eq:fg}
\end{equation}
where
we used $N_{n-1}/N_n \simeq (u+v)^{-1}$ for large $n$.
Using the expansion of the functions $G(\mu_1)$ and $h_2(\mu_1)$
given in Appendix~\ref{seca:derivation},
we can obtain the expansion
of $g(\mu_1^{(\tau)},p^{(\tau)})$ up to
the first order  in $\mu_1^{(\tau)}$ and
$p^{(\tau)}-p^*$ as
\begin{eqnarray}
&&g(\mu_1^{(\tau)},p^{(\tau)})\simeq\nonumber\\
&&{1 \over 2(u+v)(u+v-2)} \left[
\log (G(0))
+ \log\left(1-{\mu_1^{(\tau)} \over \mu_{c1}}\right)\right]\nonumber\\
&+&{1\over 2(u+v)} \left[\log[G(0)\{p^*+u^{-1}+v^{-1}\}]\right.\nonumber\\ &+&\left.
\log\left(1 +{p^{(\tau)}-p^*+\mu_1^{(\tau)} {(uv-1)(u+v)\over 3uv}
\over p^*+u^{-1}+v^{-1}}- {\mu_1^{(\tau)}\over \mu_{c1}}\right)\right],
\label{eq:gexpand}
\end{eqnarray}
where we used $L_{n-1}/N_n \simeq (u+v-1)/((u+v)(u+v-2))$ and
introduced a constant
\begin{equation}
\mu_{c1}={3\over u^2 +v^2 -2}.
\end{equation}
The renormalized values of $\mu_1$ and $p$ obtained above for
each case of (I), (II), and (III) can be used in Eqs.~(\ref{eq:fg})
and (\ref{eq:gexpand}) to evaluate the free energy $f(\mu)$.

For the case (I), the shortcuts have no weight and the network
is a fractal as long as $u\geq 2$ and $v\geq 2$.
The parameter $p$ remains at zero under the RG transformation
and $\mu_1$ is renormalized by Eq.~(\ref{eq:RG1}).
Inserting Eq.~(\ref{eq:RG1}) in Eqs.~(\ref{eq:fg})
and (\ref{eq:gexpand}) with $p^{(\tau)}=p^*=0$,
we obtain the free energy and then the spectral
density function via Eq.~(\ref{eq:rhof}), which is calculated as
\begin{eqnarray}
\rho_{\rm (I)}(\mu)&\simeq& {2\over\pi}{\rm Im} \sum_{\tau=0}^\infty
{\partial \mu_1^{(\tau)}\over \partial \mu}
{ {\partial \over \partial \mu_1^{(\tau)}} g(\mu_1^{(\tau)},0)\over (u+v)^\tau}
\nonumber\\
&\simeq& \int_{1}^\infty dz { {\delta(z-\mu_{c1}/\mu) \over (u+v)(u+v-2)} +
    {\delta(z-\mu_{c2}/\mu) \over u+v} \over \mu\log(uv) z^{\log(u+v)/\log(uv)}}
 \nonumber\\
&\simeq& \kappa_{\rm (I)}
\mu^{{\log(u+v)\over\log(uv)}-1},
\label{eq:rhomup0}
\end{eqnarray}
where
\begin{eqnarray}
\mu_{c2}=\frac{3}{u^2+v^2-uv -1}, \nonumber
\end{eqnarray}
and
\begin{eqnarray}
\kappa_{\rm (I)}=
\frac{( (u+v-2)^{-1}\mu_{c1}^{-{\log(u+v)\over\log(uv)}} +
\mu_{c2}^{-{\log(u+v)\over\log(uv)}})}{(u+v) \log(uv)}. \nonumber
\end{eqnarray}
Here we used the relation $\log x = \log |x| + i\pi \theta(-x)$ for real $x$,
where $\theta(x)$ is the Heaviside-step function.
Comparing Eq.~(\ref{eq:dsdef}) with (\ref{eq:rhomup0}),
we find that the spectral dimension $d_s$ for the case (I) is
\begin{equation}
d_s^{({\rm I})} = {2\log(u+v)\over \log(uv)}.
\label{eq:dsp0}
\end{equation}
This is consistent with the previous result obtained for the mean first passage
time studied in Ref.~\cite{rozenfeld07a}.
The spectral dimension in Eq.~(\ref{eq:dsp0}) is reduced to $2$ when
$u=v=2$, and ranges between $1$ and $2$ when $u$ and $v$ differ from $2$.
Note that it is different from the fractal dimension given in Eq.~(\ref{eq:Nlp0}).

For the case (II),  the RG flow starting from a point with non-zero values of
$\mu_1$ and $p$ first approaches the fixed point $(0,1-u^{-1}-v^{-1})$, and then it is
repelled in the direction of one eigenvector of the matrix $R$
in Eq.~(\ref{eq:RGlinear}) as shown in Fig.~\ref{fig:rg}(a).
Using Eq.~(\ref{eq:RG2}) and setting $p^* = 1-u^{-1}-v^{-1}$
in Eqs.~(\ref{eq:fg}) and (\ref{eq:gexpand}),
we obtain the spectral density function
\begin{eqnarray}
\rho_{\rm (II)}(\mu)&\simeq& {2\over\pi}{\rm Im} \sum_{\tau=0}^\infty
{\partial \over \partial \mu}
{ {\partial \over \partial \mu_1^{(\tau)}} g(\mu_1^{(\tau)},p^{(\tau)})\over (u+v)^\tau}
\nonumber\\
&\simeq& \int_{1}^\infty dz {{\delta\left(z-{\mu_{c1} \over \mu c}\right) \over (u+v)(u+v-2)} +
 {\delta\left(z-{\mu_{c3} \over \mu }\right) \over (u+v)}
\over z\mu \log(u+v)} \nonumber\\
&\simeq& \kappa_{\rm (II)},
\label{eq:rhomupgeneral}
\end{eqnarray}
where $$c= {(u+v)(uv-1)\over (uv)(u+v-1)},$$
$$\mu_{c3} = {3uv (u+v-1)\over (uv-1)(u+v)(u^2+v^2-u-v-1)},$$
and
$$
\kappa_{\rm (II)} = \frac{(uv-1)(u^2+v^2-u-v)}{3uv(u+v-2)\log(u+v)}.
$$
Thus, the spectral density function for the case (II) is a constant for small $\mu$, indicating that
\begin{equation}
d_s^{({\rm II})} = 2.
\label{eq:dsp1}
\end{equation}
This means that the spectral dimension is changed by the addition of shortcuts.
Remarkably $d_s$ remains at $2$ regardless of $u$ or $v$.

For the case (III), there is only one fixed point and
$p^{(\tau)}$ remains to be $\mathcal{O}(1)$. Inserting Eqs.~(\ref{eq:j2uv22}) and (\ref{eq:RG3}) in
Eqs.~(\ref{eq:fg}) and (\ref{eq:gexpand}) with $p^* = 0$,
the spectral density function  for $u=v=2$ is calculated as
\begin{eqnarray}
\rho_{\rm (III)}(\mu)&\simeq& {2\over\pi}{\rm Im} \sum_{\tau=0}^\infty
{\partial \mu_1^{(\tau)}\over \partial \mu}
{ {\partial \over \partial \mu_1^{(\tau)}} g(\mu_1^{(\tau)},p^{(\tau)})\over (u+v)^\tau}
\nonumber\\
&\simeq& \int_{1}^\infty {dz \over \mu z \log 4} \left[
{1\over 8} \delta\left(z-{\mu_{c1}\over \mu }{3 (1+p\log_4 z )\over p+3}\right)\right.\nonumber\\
&& \left.+{1\over 4} \delta\left(z-{1\over \mu }{3 (1+p\log_4 z )\over p+3} \right) \right]
\nonumber\\
&\simeq& {p+3\over 6 \log 4}{1\over  1+p\log_4 (1/\mu)}.
\label{eq:rhomuuv22}
\end{eqnarray}
We remark that Eq.~(\ref{eq:rhomuuv22}) with $p=0$ is reduced to
Eq.~(\ref{eq:rhomup0}).
On the other hand,   for $p>0$, the spectral density function
takes a logarithmic form,  $\rho(\mu)\sim (\log(1/\mu))^{-1}$ for
$\mu\to 0$. In the scaling regime where $\mu \exp(p^{-1})$ is finite,
the spectral density function displays the following crossover behavior:
\begin{equation}
\rho_{\rm (III)}(\mu) \simeq \left\{
\begin{array}{ll}
{p+3\over 6p} {1\over \log_4(1/\mu)} & {\rm for} \ \mu e^{p^{-1}}\ll 1,\\
{p+3\over 6 \log 4} & {\rm for} \ \mu e^{p^{-1}}\gg 1.
\end{array}
\right.
\end{equation}

\section{The Return-to-origin probability}
\label{sec:simulation}

In this section, we derive the RTO probability $P_{o}(t)$ of
RWs from the spectral density function $\rho(\mu)$
obtained in the previous section. The behavior
of $\rho(\mu)$ around $\mu=2$, which should be known for Eq.~(\ref{eq:PRTO}),
can be understood by
the symmetry property of the RG equation (\ref{eq:RGpara}).
First let us consider two cases.
(i) When both $u$ and $v$ are odd, $G(\mu_1)$  and $h_2(\mu_1)$ are symmetric
and $h_1(\mu_1)$ are anti-symmetric with respect
to $\mu_1=1$. Thus, the variable $\tilde{\mu}_1 \equiv 2-\mu_1$ and $p$ are
renormalized in the same manner as Eqs.~(\ref{eq:RG1}) and (\ref{eq:RG2})
around the fixed points $(\tilde{\mu}_1^*=2,p^*=0)$ and
$(\tilde{\mu}_1^*=2,p^*=1-u^{-1}-v^{-1})$, respectively. This leads to $\rho(2-\mu)=\rho(\mu)$.
(ii) When $p=0$ and both $u$ and $v$ are even, $G(\mu_1)$ are symmetric and
$h_1(\mu_1)$ and $h_2(\mu_1)$ are anti-symmetric with respect to $\mu_1$.
Interestingly, the parameter $\mu_1$ initially close to $2$ is renormalized
by one RG transformation to a value near zero by Eq.~(\ref{eq:RGpara}), and then
follows the same RG flow as given in Eqs.~(\ref{eq:RG1}) and (\ref{eq:RG2}).
Therefore, the spectral density function is symmetric about $\mu=1$.
In these two cases (i) and (ii), it follows that
\begin{eqnarray}
P_o(t)\simeq (1+(-1)^t)\int_0^\infty d\mu \rho(\mu)e^{-\mu t}.
\end{eqnarray}
For other cases than (i) or (ii), there is no fixed point
on the line $\mu_1=2$ and thus the contribution of $\rho(\mu)$ around $\mu=2$
to $P_o(t)$ for large $t$ is subdominant,
leading to
\begin{eqnarray}
P_o(t)\simeq \int_0^\infty d\mu \rho(\mu)e^{-\mu t}.
\end{eqnarray}

Taking account of different behaviors of $\rho(\mu)$ for small $\mu$
in the cases (I), (II), and (III), we obtain from Eq.~(\ref{eq:PRTO})
\begin{eqnarray}
P_o(t) \simeq  p_o
   \left\{
    \begin{array}{ll}
\kappa_{\rm (I)}\  \Gamma(d_s^{\rm (I)}/2) \ t^{-d_s^{\rm (I)}/2}  & {\rm (I)},  \\
\kappa_{\rm (II)} t^{-1} & {\rm (II)},\\
{p+3\over 6\log 4} t^{-1}(1+p \log_4 t)^{-1} & {\rm (III)},
\end{array}
\right.
\label{eq:RTO}
\end{eqnarray}
where $\Gamma(x)$ is the Gamma function and
the spectral dimension  $d_s^{\rm (I)}$ in the case (I) is given in Eq.~(\ref{eq:dsp0}).
Note that for the case (III), a logarithmic term appears.
As discussed above, the coefficient $p_o$ is $2$ when both $u$ and $v$ are odd or
when $p=0$ and both $u$ and $v$ are even. Otherwise, it is $1$.

\begin{figure*}
\includegraphics[width=7.5cm]{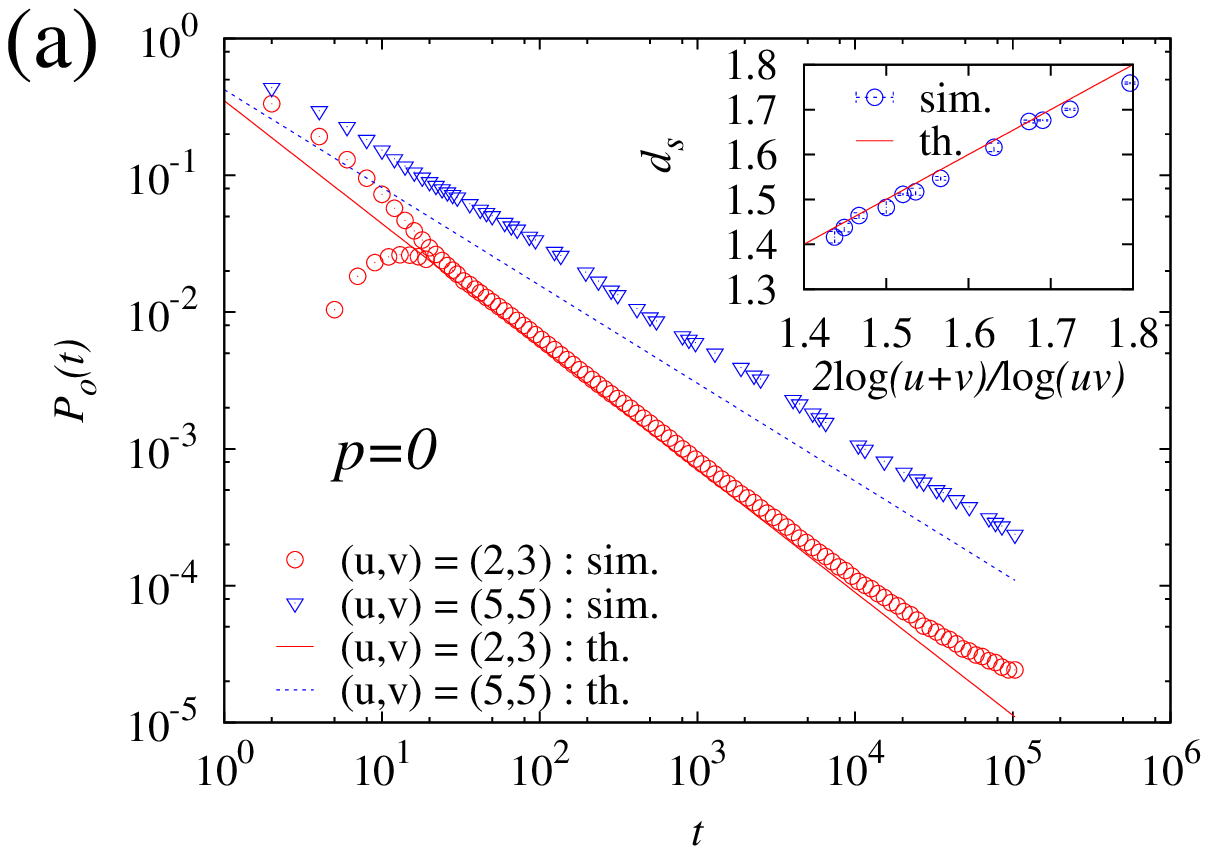}
\includegraphics[width=7.5cm]{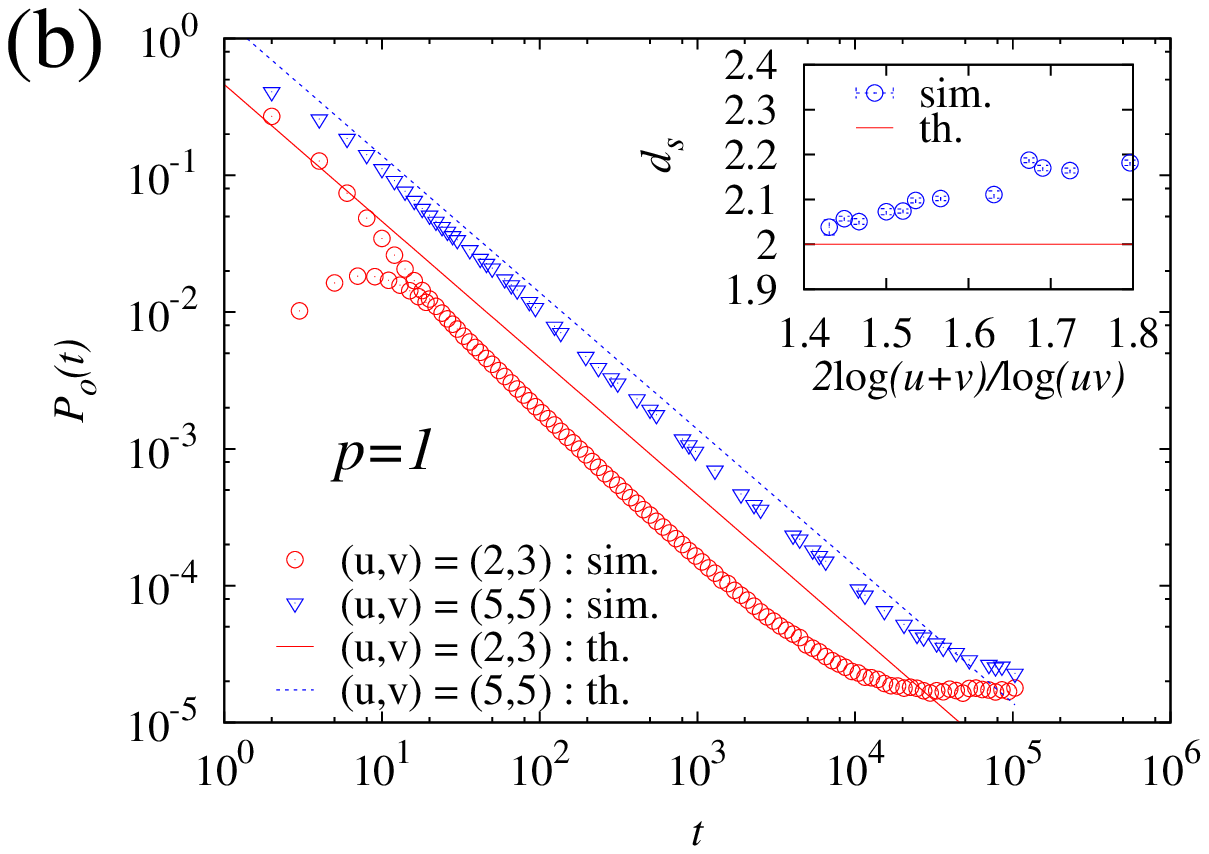}
\includegraphics[width=7.5cm]{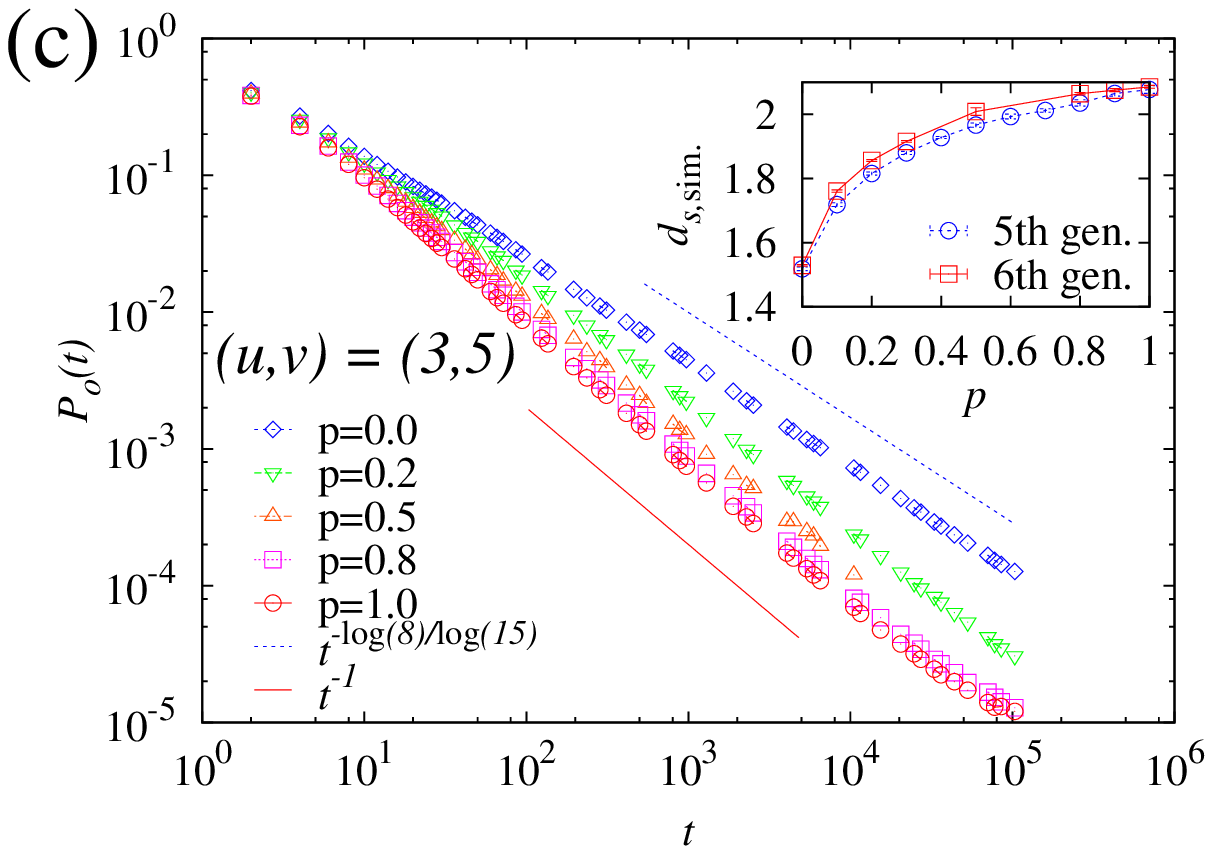}
\includegraphics[width=7.5cm]{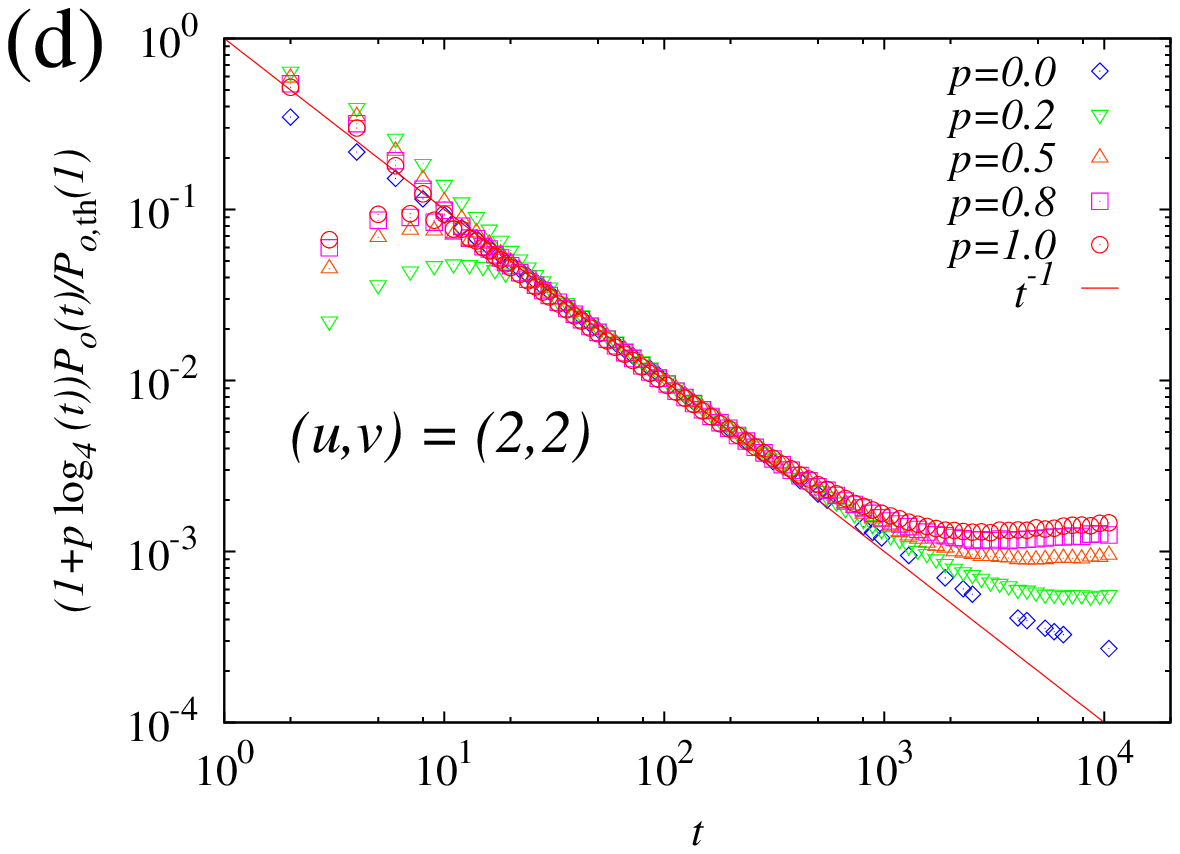}
\caption{Return-to-origin probability $P_o(t)$ on WF networks.
(a) Plots of $P_o(t)$ on the fractal WF networks ($p=0$).
Simulation results for $(u,v)=(2,3)$ and $(u,v)=(5,5)$ are shown and
the lines represent the theoretical prediction in Eq.~(\ref{eq:RTO}).
(inset) Plot of the spectral dimension $d_s$  as a function
of $2\log(u+v)/\log(uv)$. The value of $d_s$ were estimated from $P_o(t)$ for
$(u,v)=(2,3), (2,4), (2,6), (3,3), (3,4), (3,5), (3,6), (4,4), (4,5), (4,6)$, and $(5,5)$
using the relation $P_o(t)\sim t^{-d_s/2}$.
The line represents the analytic prediction $d_s = 2\log(u+v)/\log(uv)$.
(b) Plots of $P_o(t)$ on the non-fractal WF networks ($p=1$)
for the same values of $u$ and $v$.
The lines represent the theoretical predictions
in Eq.~(\ref{eq:RTO}). (inset) Plot of the spectral
dimension $d_s$ as a function of $2\log(u+v)/\log(uv)$.
The line represents the analytic prediction $d_s = 2$.
(c) Plots of $P_o(t)$ for $(u,v)=(3,5)$ and various values of $p$.
The slopes of the two lines are $-\log 8/\log 15$ and $-1$, corresponding to
the analytic prediction for $d_s/2$ for $p=0$ and $p=1$ for $(u,v)=(3,5)$.
(inset) Plot of the spectral dimension $d_s$ as a function of $p$.
The results obtained for the 5th and 6th generation are presented.
(d) Plots of $(1+p\log_4 t) P_o(t) /P_{o,{\rm th}}(1)$ versus
$t$ for $(u,v)=(2,2)$ and $p=0, 0.2, 0.5, 0.8$, and $1$, where
$P_{o,{\rm th}}(1)=p_o (p+3)/(6 \log 4)$.
All the data points collapse to $t^{-1}$ as
predicted by Eq.~(\ref{eq:RTO}).}
\label{fig:simul}
\end{figure*}

We performed numerical simulations of RWs on the WF networks.
In the simulation, $10^6$ independent RWers moved around on the networks
with the transition probability $W_{\ell q}/k_q$ in Eq.~(\ref{eq:wlq}).
The RTO probability $P_o(t)$ is measured as the fraction of the
walkers who are found in their own initial positions at $t$.
The measured RTO probabilities for various values of $p$ and ($u$,$v$)
are shown in Fig.~\ref{fig:simul}.

The spectral dimension is obtained from the power-law decay of
$P_o(t)$. When $p=0$,  the measured values of $d_s$ for various $u$ and $v$
are in good agreement with Eq.(\ref{eq:dsp0}) (see Fig.~\ref{fig:simul} (a)).
On the other hand, when $p=1$,
the numerical values of $d_s$ are somewhat larger than the theoretical
value 2, weakly dependent on $u$ and $v$ ranging $2 \sim 2.2$ as shown
in Fig.~\ref{fig:simul} (b).
Given the logarithmic term for $u=v=2$ in Eq.~(\ref{eq:RTO}),
this slight deviation as ($u,v)$ approaches $(2,2)$
can be a finite-size effect.
The faster decay of $P_o(t)$ for $p=1$ than $p=0$ is caused by shortcuts.
Other structural factors such as the degree distribution, however,
do not affect the spectral dimension, unlike the case $p=0$.

The case $0 < p < 1$ is also interesting. For $(u,v)\ne (2,2)$,
we obtain that $d_s$ increases rapidly with $p$ as shown in
the case $(u,v)=(3,5)$ in Fig.\ref{fig:simul} (c).
The curve looks more concave as $N$ increases. Thus
we expect that $d_s$ is independent of $p$ in the thermodynamic limit
as long as $p > 0$.
For $(u,v)=(2,2)$, the RTO probability is expected to have a logarithmic term as
$P_o(t)\simeq P_{o,{\rm th}}(1) t^{-1}(1+p\log_4 t)^{-1}$ with
$P_{o,{\rm th}}(1)=p_o(p+3)/(6 \log 4)$.
We plot $(1+p\log_4 t) P_o(t)/P_{o,{\rm th}}(1)$ versus $t$
for various values of $p$ in Fig.~\ref{fig:simul}(d), in which
the data are well collapsed onto $t^{-1}$ as predicted in Eq.~(\ref{eq:RTO}).

\section{Summary and discussion}
\label{sec:summary}
In this paper, we have obtained the exact solution for
the RTO probability of RWs and the spectral density function of the
Laplacian matrix in hierarchical scale-free networks by applying the
RG approach to the Gaussian model.
Particularly, we monitored the spectral dimension
as we controlled the weight of the shortcuts. When the shortcut
weight is zero,
the network is a fractal, and its spectral dimension varies
in the range $(1,2]$. It depends on the parameters
controlling network structure. On the other hand,
when the weight is larger than zero, the spectral dimension is
2, which is robust under structural variations.
This result demonstrates that shortcuts strongly affect the RW motion
even when the shortcut weight is small.

It was shown that the organization of shortcuts in a network was
essential in the classification of network structures~\cite{rozenfeld10}.
Let $p(r)$ denote the probability of finding a shortcut
between a pair of nodes that would be apart by distance $r$.
If the decay of $p(r)$ is slower than $r^{-2d_f}$ with $d_f$ being
the fractal dimension of the network, the network
is renormalized to the complete graph.
Otherwise, the network is renormalized to the corresponding fractal network.
In the non-fractal WF networks, we obtain that $p(r)\sim r^{-\alpha}$,
where $\alpha/d_f$ is in the range between $1.6$ and $2$. Thus,
the hierarchical networks with shortcuts belong to the
former case, which  can be related to our finding that
the spectral dimension $d_s$ is robust against the variation of $u$
or $v$ in the $p > 0$ regime.

We performed the numerical simulations of RWs on the
following real-world complex networks:
the protein interactions network of the yeast~\cite{yeastppi},
the human protein interaction network~\cite{humanppi},
the coauthorship network of the cond-mat archive~\cite{condmat},
and the Internet at the autonomous system level~\cite{as2004}.
Among them, the two protein interaction networks  are known as
fractals and both the coauthorship network and the Internet
are non-fractals \cite{jskim07}.
The spectral dimensions obtained from the power-law decay of the
RTO probabilities turned out to be larger for non-fractal networks than
fractal networks.
The numerical values of $d_s$ were approximately $1.4$ and $2$
for the yeast and the human protein interaction network, respectively,
and $4$ and $5.1$ for the coauthorship network and the Internet, respectively.
The probability $R$ that a RWer will ever return to the origin is related
to the RTO probability as $R=1-1/\sum_{t=0}^\infty P_o(t)$ and thus
RWs are recurrent (transient) if $d_s\leq 2 \ (d_s\geq 2)$
~\cite{hughesbook,havlinbook,rednerbook}.
In this sense, the structure of the hierarchical
networks with shortcuts studied in this work, having
$d_s=2$, is marginal in the RW motion; RWers can never return to
the origin if the network structure is more entangled.
It would be interesting to see the case of the Internet.
Since the spectral dimension is larger than $2$ for the Internet,
packets roaming on the Internet may not return to its origin or a
destination if they follow the RW motion. Therefore
the studied hierarchical network could be a good sample for
the construction of the Internet.

\begin{acknowledgments}
BK thanks Z.~Toroczkai for helpful discussions during his staying 
at the Univ. of Notre Dame. This work was supported by NRF research 
grants funded by the MEST (Nos.~2010-0015066 (BK) and 2010-0015197 (DSL)) 
and by the NAP of KRCF (DSL,DK).
\end{acknowledgments}

\appendix

\section{Other Laplacians}
\label{seca:laplacian}

The Laplacian matrix $L$, defined as
$L_{\ell q} = \delta_{\ell q} - W_{\ell q}/k_q$,
can be brought into a symmetric form by
a similarity transformation $\tilde{L} = SLS^{-1}$ with
$S_{\ell q}=k_q^{-1/2}\delta_{\ell q}$~\cite{dkim07}
\begin{equation}
\tilde{L}_{\ell q}=\delta_{\ell q}-{1\over \sqrt{k_\ell k_q}}W_{\ell q}.
\label{eq:modLaplace1}
\end{equation}
The eigenvalues of $\tilde{L}$ and $L$ are identical.
Another Laplacian matrix $L'_{\ell q}=k_\ell \delta_{\ell q}-W_{\ell q}$~\cite{dkim07} has
different eigenvalues from $L$ or $\tilde{L}$.
We consider $\tilde{L}$ in Eq.~(\ref{eq:modLaplace1}) for our analysis
of the RW problem
but the main results hold also for $L$ and $L'$~\cite{hwang_pre}.

\section{Derivation of Eq.~(\ref{eq:rgleaf})}
\label{seca:derivation}

In this section, we derive the partition function
$Z_{\rm leaf}(\phi_1',\phi_2')$ in Eq.~(\ref{eq:rgleaf}) for
a leaf in Fig.~\ref{fig:model} (b). Using Eq.~(\ref{eq:Lp1}),
we can represent $Z_{\rm leaf}(\phi_1',\phi_2')$ as
\begin{eqnarray}
&& Z_{\rm leaf} (\phi_1',\phi_2')  =
z_n^{\rm (II)}(\phi_1',\phi_2') \int \prod_{\ell=1}^{u-1} d\phi_\ell^{\rm (up)}
\prod_{q=1}^{v-1} d\phi_q^{\rm (lo)} \nonumber\\
&&\prod_{\ell=1}^{u-2} z_n^{\rm (I)}(\phi_\ell^{\rm (up)},\phi_{\ell+1}^{\rm (up)})
\prod_{q=1}^{v-2} z_n^{\rm (I)}(\phi_q^{\rm (lo)},\phi_{q+1}^{\rm (lo)})\nonumber\\
&&\times z_n^{\rm (I)}(\phi_1',\phi_{1}^{\rm (up)})
z_n^{\rm (I)}(\phi_{u-1}^{\rm (up)},\phi_2')\nonumber\\
&&\times z_n^{\rm (I)}(\phi_1',\phi_{1}^{\rm (lo)})
z_n^{\rm (I)}(\phi_{v-1}^{\rm (lo)},\phi_2') \nonumber\\
&&= z_n^{\rm (II)}(\phi_1',\phi_2') \int \prod_{\ell=1}^{u-1} d\phi_\ell^{\rm (up)}
\prod_{q=1}^{v-1} d\phi_q^{\rm (lo)} \nonumber\\
&& \exp\left[-i\sum_{\ell,q=1}^{u-1} \phi_\ell^{\rm (up)}
\bar{H}^{\rm (up)}_{\ell q} \phi_q^{\rm (up)}
-i\sum_{\ell,q=1}^{v-1} \phi_\ell^{\rm (lo)} \bar{H}^{\rm (lo)}_{\ell q} \phi_q^{\rm (lo)}\right.
 \nonumber\\
&& +2i \phi_1' (\phi_1^{\rm (up)} + \phi_1^{\rm (lo)}) +
2i \phi_2' (\phi_{u-1}^{\rm (up)} + \phi_{v-1}^{\rm (lo)}) \nonumber\\
&& \left.-2i(1-\mu_1)(\phi_1^{'2} + \phi_2^{'2})\right].
\label{eqa:rgleaf}
\end{eqnarray}
Here we introduced an $(u-1)\times (u-1)$ Hamiltonian
$\bar{H}^{\rm (up)}$ and a $(v-1)\times (v-1)$
Hamiltonian $\bar{H}^{\rm (lo)}$, both of which take a form
\begin{equation}
\bar{H}=\left(
    \begin{array}{ccccccc}
    H_0 & -1 & 0 & \cdots & \cdots & \cdots& 0\\
    -1 & H_0 & -1 & 0 & \cdots & \cdots& 0\\
    0 &-1 & H_0 & -1 & 0 & \cdots& 0\\
    \cdots && \cdots && \cdots && \cdots \\
    0 & \cdots &\cdots & \cdots & 0 & -1 & H_0
    \end{array}
    \right)
\end{equation}
with $H_0 = 2(1-\mu_1)$. The $(n-1)\times (n-1)$ matrices in this form
have $n-1$ eigenvalues $\lambda_\ell$ and the corresponding eigenvectors ${\bf e}_\ell$
with $\ell=1,2,\ldots,n-1$~\cite{chengbook}
\begin{eqnarray}
\lambda_\ell&=&H_0 - 2\cos\left({\ell\pi\over n}\right),\nonumber\\
{\bf e}_\ell&=&\sqrt{2\over n}\left(
    \sin\left({\ell\pi\over n}\right),
    \sin\left({2\ell\pi\over n}\right),
    \ldots\right.\nonumber\\
&& \left.
\ldots,\sin\left({(n-1)\ell\pi\over n}\right)
\right).
\label{eqa:lambda}
\end{eqnarray}
Therefore $\bar{H}^{\rm (up)}$ and $\bar{H}^{\rm (lo)}$ are
diagonalized with bases
\begin{eqnarray}
\tilde{\phi}_\ell^{\rm (up)} &=& \sqrt{2\over u} \sum_{q=1}^{u-1} \sin\left({\ell q\pi\over u}\right)\phi_q^{\rm (up)}, \ {\rm and}\nonumber\\
\tilde{\phi}_\ell^{\rm (lo)} &=& \sqrt{2\over v} \sum_{q=1}^{v-1} \sin\left({\ell q\pi\over v}\right)\phi_q^{\rm (lo)},
\end{eqnarray}
respectively,
and $Z_{\rm leaf}(\phi_1',\phi_2')$ is rewritten as
\begin{eqnarray}
&&Z_{\rm leaf} (\phi_1',\phi_2')  =
\exp\left[-i(1-\mu_2)p (\phi_1'^2 +\phi_2'^2)
    \right.\nonumber\\
&& \left. +2i p \phi_1' \phi_2'\right]
 \int \prod_{\ell=1}^{u-1} d\tilde{\phi}_\ell^{\rm (up)}
\prod_{q=1}^{v-1} d\tilde{\phi}_q^{\rm (lo)}
\exp\left[ \right.\nonumber\\
&& -i \sum_{\ell=1}^{u-1} \lambda_\ell (\tilde{\phi}_\ell^{\rm (up)})^2
-i \sum_{q=1}^{v-1} \lambda_q (\tilde{\phi}_q^{\rm (lo)})^2
+2i(\phi_1' +\phi_2')\left\{ \right.\nonumber\\
&&\left.\sqrt{2\over u} \sum_{\ell=1}^{u-1} \sin\left({\ell\pi\over u}\right) \tilde{\phi}_\ell^{\rm (up)}  +
\sqrt{2\over v} \sum_{q=1}^{v-1} \sin\left({q\pi\over v}\right) \tilde{\phi}_q^{\rm (lo)}
\right\} \nonumber\\
&&\left. -2i(1-\mu_1)(\phi_1'^2 +\phi_2'^2)\right],
\label{eqa:rgleaf2}
\end{eqnarray}
which is calculated to give
Eq.~(\ref{eq:rgleaf}). Note that $c(\ell,a,b)$ in
Eq.~(\ref{eq:ch1h2}) is
half the eigenvalue $\lambda_\ell$ in
Eq.~(\ref{eqa:lambda}) with $n=b$ and $H_0=2(1-a)$.

\section{Expansion of $G(\mu_1), h_1(\mu_1)$, and $h_2(\mu_1)$ in Eq.~(\ref{eq:ch1h2})}
\label{seca:expansion}

The small-$\mu_1$ behaviors of $G(\mu_1),
h_1(\mu_1)$, and $h_2(\mu_1)$ are used to linearize
Eq.~(\ref{eq:RGpara}) and to find the small-$\mu_1$
behaviors of $g(\mu_1,p)$ in Eq.~(\ref{eq:gmu}).
Expanding these functions for small $\mu_1$ up to the first order,
we obtain
\begin{eqnarray}
G(\mu_1) &=& G(0)\left[1 -{\mu_1\over 2} \{S_1 (u) + S_1 (v)\}\right] + \mathcal {O}(\mu_1^2),\\
h_1 (\mu_1) &=& {u-1\over u}+{v-1\over v} + \mu_1 \times\nonumber\\
   && \left[{S_1(u)-(u-1) \over u} +
{S_1(v)-(v-1)\over v}\right]  + \mathcal {O}(\mu_1^2), \nonumber\\
h_2 (\mu_1) &=& {1\over u}+{1\over v} + \mu_1 \left[{S_2(u)-{1+(-1)^u\over 2}\over u} \right.\nonumber\\
&&\left.+{S_2(v)-{1+(-1)^v\over 2}\over v}\right] + \mathcal {O}(\mu_1^2),
\label{eqa:Gh1h2}
\end{eqnarray}
where $G(0)=\prod_{\ell=1}^{u-1} (1-\cos(\ell\pi/u))\prod_{q=1}^{v-1}(1-\cos(q\pi/v))$ and we
introduced two sums $S_1(n)$ and $S_2(n)$ defined as
\begin{eqnarray}
S_1(n) & = & \sum_{j=1}^{n-1} \frac{1}{\sin^2(\frac{j \pi}{2n})}, \nonumber \\
S_2(n) & = & \sum_{j=1}^{n-1} \frac{(-1)^{j+1}}{\sin^2(\frac{j \pi}{2n})}.
\end{eqnarray}
To compute $S_1(n)$ and $S_2(n)$, we use the following identity
\begin{equation}
{1\over \sin^2 (\pi a)} = {1\over \pi^2}
\sum_{k=-\infty}^{\infty} {1\over (k-a)^2},
\label{eqa:id}
\end{equation}
which can be derived by considering the following
integral on a circle the radius $R$ of which goes to infinity:
\begin{equation}
\int dz {\cot (\pi z )\over (z-a)^2} = i
\left\{\sum_{n=-\infty}^\infty {1\over (n-a)^2 } - {\pi^2 \over \sin^2 (\pi a)}\right\}=0.
\end{equation}
Using Eq.~(\ref{eqa:id}), the sum $S_1(n)$ is evaluated as
\begin{eqnarray}
S_1(n) & = & \sum_{j=1}^{n-1} \frac{1}{\sin^2(\frac{j \pi}{2n})} \nonumber \\
& = & \frac{1}{2} \sum_{j=1}^{n-1} \left(\frac{1}{\sin^2(\frac{j \pi}{2n})} + \frac{1}{\sin^2(\pi-\frac{j \pi}{2n})}\right) \nonumber \\
& = & \frac{1}{2 \pi^2} \sum_{j=1}^{n-1} \sum_{k=-\infty}^{\infty}
\left(\frac{1}{(\frac{j}{2n}-k)^2} + \frac{1}{(1-\frac{j}{2n}-k)^2}\right) \nonumber \\
& = & \frac{4 n^2}{\pi^2}\sum_{s=1}^{\infty} \left(\frac{1}{s^2} - \frac{1}{(sn)^2}\right) = \frac{2(n^2-1)}{3}.
\label{eqa:S1}
\end{eqnarray}
Similarly, the sum $S_2(n)$ is given by
\begin{eqnarray}
S_2(n) & = & \sum_{j=1}^{n-1} \frac{(-1)^{j+1}}{\sin^2(\frac{j \pi}{2n})} \nonumber \\
& = & \frac{4 n^2}{\pi^2} \sum_{j=1}^{n-1} \sum_{k=0}^{\infty} \frac{(-1)^{j+1}}{(j-2nk)^2} + \frac{(-1)^{j+1}}{(-j-2n(k-1))^2} \nonumber \\
& = & \left\{ \begin{array}{cc} \frac{4 n^2}{\pi^2}\sum_{s=1}^{\infty} \frac{(-1)^{s+1}}{s^2} + \frac{1}{(sn)^2} = \frac{n^2+2}{3} & \mbox{n is even},\\
\frac{4 n^2}{\pi^2}\sum_{s=1}^{\infty} \frac{(-1)^{s+1}}{s^2} - \frac{(-1)^{j+1}}{(sn)^2} = \frac{n^2-1}{3} & \mbox{n is odd},
\end{array} \right.
 \label{B}
 \nonumber\\
&=&{n^2-1\over 3} + {1+(-1)^n\over 2}.
\label{eqa:S2}
\end{eqnarray}

Inserting Eqs.~(\ref{eqa:S1}) and (\ref{eqa:S2})
into Eq.~(\ref{eqa:Gh1h2}), we find that
\begin{eqnarray}
G(\mu_1) &\simeq& G(0)\left(1 -\mu_1 {u^2+v^2-2\over 3}\right),\nonumber\\
h_1 (\mu_1) &\simeq& 2 - {u+v \over uv} + \mu_1 \times\nonumber\\
 && \left({2u^2-3u+1 \over 3u} +
    {2v^2-3v+1\over 3v}\right), \nonumber\\
h_2 (\mu_1) &\simeq& {u+v\over uv} + \mu_1 \left({u^2-1 \over 3u} + {v^2-1\over 3v}\right).
\label{eqa:Gh1h2expand}
\end{eqnarray}


\begin{references}
\bibitem{hughesbook} B.D. Huges, {\it Random Walks and Random Environments}
(Oxford Univ. Press, Clarendon, 1995).
\bibitem{havlinbook} D. ben-Avraham and S. Havlin, {\it Diffusion and Reactions in
  Fractals and Disordered Systems} (Cambridge Univ. Press, Cambridge, 2000).
\bibitem{rednerbook} S. Redner, {\it A Guide to First-Passage Processes} (Cambridge Univ. Press, Cambridge, 2001).
\bibitem{albert02} R. Albert and A.-L. Barab\'asi, Rev. Mod. Phys. {\bf 74,} 47 (2002).
\bibitem{dorogovtsev03a} S.N dorogovtsev and J.F.F. Mendes, Adv. Phys. {\bf 51}, 1079 (2002).
\bibitem{newman03} M.E.J. Newman, SIAM Rev. {\bf 45}, 167 (2003).
\bibitem{sven01} S. Bilke and C. Peterson, Phys. Rev. E {\bf 64}, 036106 (2001).
\bibitem{noh04} J.D. Noh and H. Rieger, Phys. Rev. Lett. {\bf 92}, 118701 (2004).
\bibitem{masuda04} N. Masuda and N. Konno, Phys. Rev. E {\bf 69}, 066113 (2004).
\bibitem{sood05} V. Sood, S. Redner, and D. ben-Avraham, J. Phys. A: Math. Gen. {\bf 38}, 109 (2005).
\bibitem{bollt05} E.M. Bollt and D. ben-Avraham, New J. Phys. {\bf 7}, 26 (2005).
\bibitem{condamin07} S. condamin, O. B\'{e}nichou, V. Tejedor, R. Voituriez, and J. Klafter, Nature {\bf 450}, 77 (2007).
\bibitem{baronchelli08} A. Baronchelli, M. Catanzaro, and R. Pastor-Satorras, Phys. Rev. E {\bf 78}, 011114 (2008).
\bibitem{zhang09} Z. Zhang, Y. Qi, S. Zhou,  W. Xie, and J. Guan, Phys. Rev. E {\bf 79}, 021127 (2009).
\bibitem{tejedor09} V. Tejedor, O. B\'{e}nichou, and R. Voituriez, Rev. E {\bf 80}, 065104(R) (2009).

\bibitem{dhkim04} D.-H. Kim, J.-D. Noh, and H. Jeong, Phys. Rev. E {\bf 70}, 046126 (2004).
\bibitem{goh06} K.-I. Goh, G. Salvi, B. Kahng, and D. Kim, Phys. Rev. Lett. {\bf 96}, 018710 (2006).
\bibitem{wu06} Z. Wu, L.A. Braunstein, S. Havlin, and H.E. Stanley, Phys. Rev. Lett. {\bf 96}, 148702 (2006).
\bibitem{rozenfeld10} H.D. Rozenfeld, C. Song, and H.A. Makse, Phys. Rev. Lett. {\bf 104}, 025701 (2010).
\bibitem{gaussian} D. Kim, J. Kor. Phys. Soc {\bf 17}, 3 (1984); R. Burioni and D. Cassi, Phys. Rev. Lett. {\bf 76}, 1091 (1996); K. Hattori, T. Hattori, and H. Watanabe, Prog. Theor. Phys. Suppl. {\bf 92}, 108 (1987).
\bibitem{berker79} A.N. Berker and S. Ostlund, J. Phys. C {\bf 12}, 4961 (1979).
\bibitem{berker06} M. Hinczewski and N. Berker, Phys. Rev. E {\bf 73}, 066126 (2006).
\bibitem{rozenfeld07a} H. D. Rozenfeld, S. Havlin and D. ben-Avraham, New J. Phys. {\bf 9}, 175 (2007).
\bibitem{rozenfeld07b} H.D. Rozenfeld and D. ben-Avraham, Phys. Rev. E {\bf 75}, 061102 (2007).
\bibitem{noh02} J.D. Noh and H. Rieger, Phys. Rev. E {\bf 66}, 066127 (2002).
\bibitem{song06} C. Song, S. Havlin, and H. A. Makse, Nature {\bf 433}, 392 (2005); Nature Physics {\bf 2}, 275 (2006).
\bibitem{shortestpath}
In Ref.~\cite{berker06}, Eq.~(\ref{eq:Nlp1}) was analytically derived for $u=v=2$ and $p=0$.
Also our numerical computation of the mean distance using the link cost $c_{\ell q}=1/W_{\ell q}$
confirms the logarithmic scaling of the mean distance in the WF networks with $p>0$.
\bibitem{perron} The eigenvalues of $L$ are distributed between $0$ and $2$ as stated in the Perron-Frobenius theorem.
\bibitem{cardybook} J. Cardy, {\it Scaling and Renormalization in Statistical Physics} (Cambridge Univ. Press, Cambridge, 1996).
\bibitem{yeastppi} J.-D. Han et al., Nature {\bf 430}, 88 (2004).
\bibitem{humanppi} Database of Interacting Proteins, http://dip.doe-mbi.ucla.edu/dip/.
\bibitem{condmat} M. E. J. Newman, Proc. Natl. Acad. Sci. U.S.A. {\bf 98}, 404 (2001).
\bibitem{as2004} University of Oregon Route Views Archive Project, http://archive.routeviews.org/.
\bibitem{jskim07} J.S. Kim, K.-I. Goh, G. Salvi, E. Oh, B. Kahng and D. Kim, Phys. Rev. E {\bf 75}, 016110 (2007).
\bibitem{dkim07} D. Kim and B. Kahng, Chaos {\bf 17}, 026115 (2007).
\bibitem{hwang_pre} S. Hwang, D.-S. Lee, B. Kahng, and D. Kim, unpublished (2010).
\bibitem{chengbook} S.S. Cheng, {\it Partial Difference Equations}  (Taylor \& Francis, London, 2003).
\end{references}
\end{document}